\documentclass[aip,reprint,rsi]{revtex4-1}
\usepackage{graphicx}

\begin{document}

\title{Real-time monitoring via second-harmonic interferometry of a flow gas cell for laser wakefield acceleration} 

\author{F. Brandi}
\email{fernando.brandi@ino.it}
\affiliation{Intense Laser Irradiation Laboratory (ILIL), Istituto Nazionale di Ottica (INO-CNR), Via Moruzzi 1, 56124 Pisa, Italy}
\affiliation{Istituto Italiano di Tecnologia (IIT), Via Morego 30, 16163 Genova, Italy}

\author{F. Giammanco}
\author{F. Conti}
\affiliation{Dipartimento di Fisica, Universit\`a degli Studi di Pisa, Largo B. Pontecorvo 3, 56127 Pisa, Italy}
\affiliation{Plasma Diagnostics \& Technologies Ltd., via Matteucci n.38/D, 56124 Pisa, Italy}

\author{F. Sylla}
\affiliation{SourceLAB SAS, 86 rue de Paris, 91400 Orsay, France}

\author{G. Lambert}
\affiliation{LOA,  ENSTA  ParisTech,  CNRS, Ecole  Polytechnique,  Université  Paris-Saclay,
828  bd  des  Maréchaux,  91762  Palaiseau  Cedex,  France}

\author{L. A. Gizzi}
\affiliation{Intense Laser Irradiation Laboratory (ILIL), Istituto Nazionale di Ottica (INO-CNR), Via Moruzzi 1, 56124 Pisa, Italy}

\date{\today}

\begin{abstract}

The use of a gas cell as a target for laser weakfield acceleration (LWFA) offers the possibility to obtain stable and manageable laser-plasma interaction process, a mandatory condition for practical applications of this emerging technique, especially in multi-stage accelerators.
In order to obtain full control of the gas particle number density in the interaction region, thus allowing for a long term stable and manageable LWFA, real-time monitoring is necessary. In fact, the ideal gas law cannot be used to estimate the particle density inside the flow cell based on the preset backing pressure and the room temperature
because the gas flow depends on several factors like tubing, regulators and valves in the gas supply system, as well as vacuum chamber volume and vacuum pump speed/throughput.
Here, second-harmonic interferometry is applied to measure the particle number density inside a flow gas cell designed for LWFA. 
The results demonstrate that real-time monitoring is achieved, and that using low backing pressure gas ($<1$ bar) and different cell orifice diameters ($<$ 2 mm) it is possible to finely tune the number density up to the $10^{19}$ cm$^{-3}$ range well suited for LWFA.

\end{abstract}

\maketitle 

After several decades of fundamental research on high-intensity laser-plasma interaction, recent progress on the production of energetic particle beams from laser wakefield acceleration (LWFA) \cite{Esarey2009} has opened the way for actual applications in future accelerator technology \cite{SchroederPRSTAB2010,MaierPRX2012} as well as for medical uses including therapy and diagnosis \cite{MalkaMed2010,GlinecMed2006,ColeSR2015,ilil2015,}. To implement these practical applications full control of the LWFA process is mandatory. A key role in LWFA is played by the free-electron density in the plasma, which is directly related to the particle number density $N$. 
Different kind of gaseous targets can be employed in LWFA. {\it Supersonic gas jets} are widely used as they allow for generating a flat density profile bounded by steep gradients. Density profiles can however vary shot-to-shot due to several reasons, including reproducibility of valve operation over time and flow turbulences. Moreover, high repetition rate operation is limited by valve pulsing capability. {\it Capillary discharges} are good candidates for LWFA, as the laser can be guided over a few centimeters, thus increasing the acceleration length and the final energy of electrons \cite{Leemans2014}. These targets are yet difficult to setup, get damaged by repetitive use, and require sophisticated characterization methods \cite{dani2015} which hampers real-time monitoring. Using {\it gas cell targets} is preferable to avoid the above mentioned limitations, enabling a stable and manageable laser-plasma interaction process, even with high repetition rate laser, along with easily tunable 
accelerator length in order to vary electron energy \cite{Osterhoff2008,corde13}. 
Moreover, flow gas cells are ideal candidates to be implemented in multi-stage accelerators \cite{poll11,jsli11}. All these features are extremely important in the perspective of designing and implementing LWFA-based facilities with superior beam quality and reliability necessary for actual high-level applications \cite{eupraxia}.  
However, $N$ in a flow gas cannot be inferred from the backing pressure and ambient temperature using the ideal gas law \cite{jju2012}, thus real-time monitor is needed.
$N$ can be accurately measured by interferometric techniques. Standard two-arm interferometers (e.g., Mach-Zehnder interferometer) are widely used for the characterization of gas targets \cite{gizz06}, but suffer from a high sensitivity on the environmental conditions. The implementation of two-arm interferometers with a quasi-common-path configuration (e.g., Nomarski interferometer) reduces the influence of environmental conditions, but requires substantial data manipulation and analysis to extract the actual density from the recorded interferograms \cite{gizz94}, limiting their applicability for a real-time measurement. For the above mentioned reasons, the use of standard interferometric approaches for an accurate real-time monitoring over an extended period of time is prevented in practice. 
An alternative robust and fast method to precisely measure $N$ is based on the so-called second-harmonic interferometer
(SHI) \cite{hopf80,drac93} a single-arm, two-color interferometer, 
with a measured phase shift given by $\Delta\phi=\frac{4\pi}{\lambda}\int_L \Delta n(\lambda)dl=\frac{4\pi}{\lambda}L \overline{\Delta n}(\lambda)$, where
$\Delta n(\lambda) = n(\lambda)-n(\lambda/2)$,  $n(\lambda)$ is the refractive index, $\lambda$ is the wavelength, and $L$ the optical path in the sample.
As a comparison, in two-arm interferometers the measured phase shift is given by $\frac{2\pi}{\lambda}\int_L (n(\lambda)-1)dl$. The value of the refractivity 
$(n(\lambda)-1)$ for gases in the visible-NIR range is  $\sim$2 orders of magnitude larger then the corresponding $\Delta n(\lambda)$. Therefore,
the lengthy data analysis and reduced accuracy related to fringe jump \cite{juhn,jju2012} when measuring large density variations with two-arm interferometers may be avoided using the SHI. As example, for a path length of 50 mm in argon at standard temperature and pressure ($\Delta n_0($1064 nm$) = 40  \times 10^{-7}$) \cite{vels86} the phase shift experienced by the SHI is 2.36 rad, while the corresponding phase shift in a two-arm interferometer is 76.4 rad \cite{bide81}.\\ 
A high-speed ($\sim \mu$s) and high-sensitivity ($\sim$ mrad) SHI employing a CW Nd:YAG laser and phase detection has been developed \cite{bran07} and implemented to measure electron density in a large plasma \cite{bran09}, number density in a pulsed gas jet \cite{bran11}, and to perform quantitative phase dispersion imaging \cite{bran13}. Here, we report on the use of such a compact SHI to measure $N$ in real-time inside a flow gas cell of variable length designed specifically for LWFA experiments. In fact, the SHI satisfies the requirements for a fast and reliable measurement, providing an efficient method to control $N$ within the interaction region. \\ 
 \begin{figure}[!t]
\centering
\includegraphics[width=\linewidth]{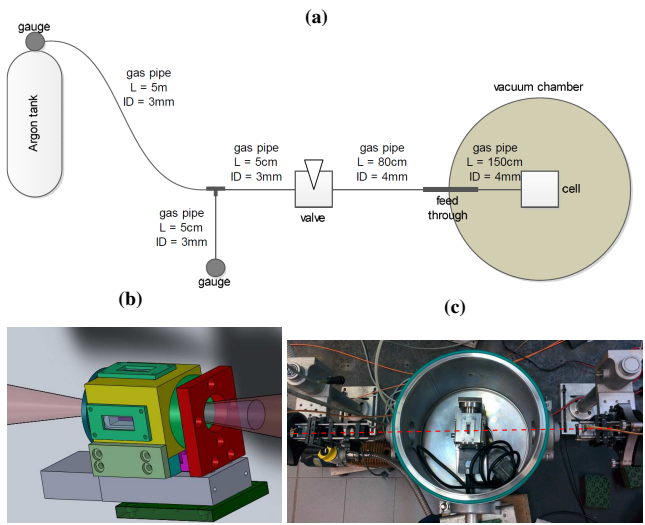}
\caption{(a) schematic of the gas supply system: L-pipe length, ID-pipe internal diameter; (b) schematic of the flow gas cell assembly; 
(c) photo of the cell in the vacuum chamber with the interferometer optical path highlighted by red dashed line.}
\label{fig1}
\end{figure}
\begin{figure}[!t]
\centering
\includegraphics[width=0.91\linewidth]{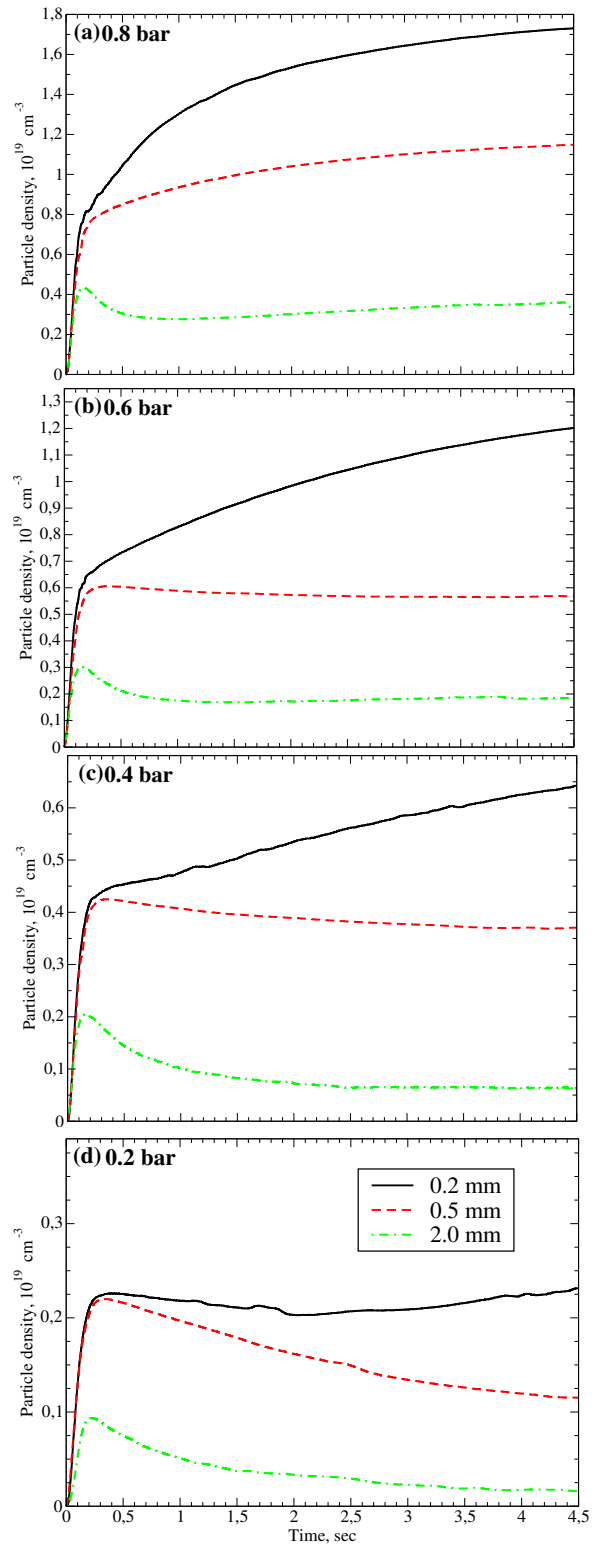}
\caption{\label{fig2} The evolution of $\overline{N}$ in the cell for different backing pressure and orifice aperture sizes: 
(a) 0.8 bar; (b) 0.6 bar; (c) 0.4 bar; (d) 0.2 bar. In all figures the continuous (black), dashed (red) and dot-dash (green) lines 
correspond to 0.2 mm, 0.5 mm and 2 mm orifice diameters respectively.}
\end{figure}
The gas system set up used in the experiment is schematically shown in Fig. \ref{fig1}(a). The gas flow cell is placed inside a vacuum chamber 
that comprises optical windows used to couple the laser light in and out, a feed through for the gas pipe, and a scroll vacuum pump.   
The gas is supplied to the cell by means of an electronic valve, and the applied backing pressure is set by the gas cylinder regulator and measured by a gauge with 50 mbar resolution placed in proximity to the valve. 
A 3D schematic of the gas cell is reported in Fig. \ref{fig1}(b). The cell comprises two lateral glass windows that allow transverse optical access for the interferometric measurements ($L=$50 mm), 
and two longitudinal orifice apertures to couple in and out the high-power ultrashort laser pulse and the produced particle beam respectively.
The two orifices, which in the present experiment are set 12 mm apart, allow also the gas to flow from the cell into the vacuum chamber. Three pairs of orifices are used, with different aperture diameters, namely 0.2 mm, 0.5 mm and 2 mm. 
In Fig. \ref{fig1}(c) a photograph of the apparatus comprising the vacuum chamber, the cell and the SHI is shown, with the optical path of the interferometer highlighted by the dashed line. 
The quantity acquired by the SHI is given by $V\sin(\Delta \phi + \phi_0) +\alpha$,  \cite{bran07} where $\phi_0$ is the off-set phase that can be reduced below 1 mrad acting on a phase compensator, $V$ is the fringe visibility, and $\alpha \ll 1$ is related to the detector responsivities. The visibility is directly obtained by scanning the phase compensator over half-fringe \cite{bran08,bran08a} and for the present experiment is found to be $V=0.9$.  

In general the refractivity $(n-1)$ of a gas is related to $N$ which is determined by the equation of state at a given temperature $T$ and pressure $P$. In the present case the Gladston-Dale relation between the refractivity and the number density, $(n-1) \propto N$, and ideal gas equation of state, $N \propto P/T$, are assumed \cite{malk00}. Therefore, the average particle number density is related to the measured phase by the equation 
$\overline{N} = \frac{\lambda}{4\pi L}\frac{N_0}{\Delta n_0} \Delta \phi=1.14 \Delta \phi \times 10^{19}$ cm$^{-3}$, 
where $N_0=2.69 \times 10^{19}$ cm$^{-3}$ is the Loschmidt constant. \\
The results of the systematic measurements for four values of the backing pressure, i.e., 0.2, 0.4, 0.6, and 0.8 bar are reported in Fig. \ref{fig2}.

The actual gas flow can be described as follows: i) the filling up of the cell starts at the backing pressure given by the preset value; ii) after the first 100 ms, the orifices start to play a role and the filling up rate decreases with increasing aperture diameter; iii) when the gas flow demand exceeds the controlling capabilities of the regulator the pressure at the cell inlet drops, reducing the cell's filling up rate; iv) in case of large apertures and/or low preset backing pressure 
$N$ decreases at longer time due to the higher gas flow from the cell to the vacuum chamber compared to the flow from the gas supply system to the cell. \\
In conclusion, it is demostrated that the SHI can be used to monitor $N$ in real-time and it is found that  the ideal gas law cannot indeed be used to estimate $N$ inside the flow cell solely based on the preset backing pressure and the room temperature, i.e., $\sim 2.5\times 10^{19}$cm$^{-3}$bar$^{-1}$. This is because the actual gas flow depends on several factors like tubing, regulators and valves in the gas supply system, as well as vacuum chamber volume, vacuum pump speed/throughput, and cell's orifice diameters.   
In fact, for the same backing pressure $N$ can differ by almost a factor of 2 when using 0.2 mm or 0.5 mm orifice. Moreover, in a repetitive operation, changes of the orifice's diameter in time due to laser ablation should be taken into account,  
thus confirming that a real-time monitoring is required to maintain a long-term stable $N$ inside the cell. 

FB and LG received funding from the European Union’s Horizon 2020 research and innovation programme under Grant Agreement No 653782 – EuPRAXIA, and from the Istituto Italiano di Tecnologia (convenzione operativa IIT INO-CNR prot. n. 0010983, 26/11/2013). The Authors acknowledge
Emilien Guillaume, C\'{e}dric Thaury for useful discussions.


\begin{thebibliography}{99}

\bibitem{Esarey2009}
E. Esarey, C. B. Schroeder, and W. P. Leemans
 Rev. Mod. Phys. \textbf{81}, 1229 (2009).

\bibitem{SchroederPRSTAB2010}
C. B. Schroeder, E. Esarey, C. G. R. Geddes, C. Benedetti, and W. P. Leemans
 Phys. Rev. ST Accel. Beams \textbf{13}, 101301 (2010).

\bibitem{MaierPRX2012}
A. R. Maier, A. Meseck, S. Reiche, C. B. Schroeder, T. Seggebrock, and F. Grüner 
 Phys. Rev. X \textbf{2}, 031019 (2012).



\bibitem{MalkaMed2010}
V. Malka, J. Faure, and Y. A. Gauduel
Mutat. Res.-Rev. Mutat. \textbf{704}, 142 – 151 (2010)

\bibitem{GlinecMed2006}
Y. Glinec, J. Faure, V. Malka, T. Fuchs, H. Szymanowski, and U. Oelfke 
Med. Phys. \textbf{33}, 155–162 (2006)

\bibitem{ColeSR2015}
J. M. Cole, J. C. Wood, N. C. Lopes, K. Poder, R. L. Abel, S. Alatabi, J. S. J. Bryant, A. Jin, S. Kneip, K. Mecseki, D. R. Symes, S. P. D. Mangles  and Z. Najmudin. 
 Sci. Rep. \textbf{5},    13244 (2015).
 
 \bibitem{ilil2015}
L.A. Gizzi, L. Labate, F. Baffigi, F. Brandi, G.C. Bussolino, L. Fulgentini, P. Koester, D. Palla, and F. Rossi,
Nucl. Instr. Meth. Phys. Res. B \textbf{355}, 241-245 (2015).

\bibitem{Leemans2014}
W. Leemans, A. Gonsalves, H.-S. Mao, K. Nakamura, C. Benedetti, C. Schroeder, C. T\'{o}th, J. Daniels, D. Mittelberger, S. Bulanov, J.-L. Vay, C. Geddes,  and E. Esarey 
Phys. Rev. Lett. \textbf{113}, 245002 (2014).


\bibitem{dani2015}
J. Daniels, J. van Tilborg, A. J. Gonsalves, C. B. Schroeder, C. Benedetti, E. Esarey, and W. P. Leemans
Phys. Plasmas \textbf{22}, 073112 (2015).

\bibitem{Osterhoff2008}
J. Osterhoff, A. Popp, Zs. Major, B. Marx, T. P. Rowlands-Rees, M. Fuchs, M. Geissler, R. Hörlein, B. Hidding, S. Becker, E. A. Peralta, U. Schramm, F. Grüner, D. Habs, F. Krausz, S. M. Hooker, and S. Karsch 
 Phys. Rev. Lett. \textbf{101}, 085002 (2008).

\bibitem{corde13}
S. Corde, C. Thaury, A. Lifschitz, G. Lambert, K. Ta Phuoc, X. Davoine, R. Lehe, R. D. Douillet, A. Rousse and V. Malka
Nature Comm. \textbf{4}, 1501 (2013).

\bibitem{poll11}
B. B. Pollock, C. E. Clayton, J. E. Ralph, F. Albert, A. Davidson, L. Divol, C. Filip, S. H. Glenzer, K. Herpoldt, W. Lu, K. A. Marsh, J. Meinecke, W. B. Mori, A. Pak, T. C. Rensink, J. S. Ross, J. Shaw, G. R. Tynan, C. Joshi, and D. H. Froula
Phys. Rev. Lett. \textbf{107}, 045001 (2011).

\bibitem{jsli11}
J. S. Liu, C. Q. Xia, W. T. Wang, H. Y. Lu, Ch. Wang, A. H. Deng, W. T. Li, H. Zhang, X. Y. Liang, Y. X. Leng, X. M. Lu, C. Wang, J. Z. Wang, K. Nakajima, R. X. Li, and Z. Z. Xu
Phys. Rev. Lett. \textbf{107}, 035001 (2011). 

\bibitem{eupraxia}
http://www.eupraxia-project.eu/

\bibitem{jju2012}
J. Ju and B. Cros
J. Appl. Phys. \textbf{112}, 113102 (2012).

\bibitem{gizz06}
L. A. Gizzi, M. Galimberti, A. Giulietti, D. Giulietti, P. Koester, L. Labate, P. Tomassini, Ph. Martin, T. Ceccotti, P. De Oliveira, and P. Monot
Phys. Rev. E \textbf{74}, 036403 (2006).

\bibitem{gizz94}
L. A. Gizzi, D. Giulietti, A. Giulietti, T. Afshar-Rad, V. Biancalana, P. Chessa, C. Danson, E. Schifano, S. M. Viana, and O. Willi
Phys. Rev. E \textbf{49}, 5628--5643 (1994).

 \bibitem{hopf80}
F.~Hopf, A.~Tomita, and G.~Al-Jumaily, 
Opt. Lett. \textbf{5}, 386--388 (1980).

  
\bibitem{drac93}
V. P. Drachev, Yu. I. Krasnikov, and P. A. Bagryansky,
Rev. Sci. Instrum. \textbf{64}, 1010--1013 (1993).

  
\bibitem{juhn} 
J.-W. Juhn, K. C. Lee, Y. S. Hwang, C. W. Domier, N. C. Luhmann Jr., B. P. Leblanc, D. Mueller, D. A. Gates,
and R. Kaita,
Rev. Sci. Instrum. \textbf{81}, 10D540 (2010).
  
   \bibitem{vels86}
 S.P. Velsko and D. Eimerl
 Appl. Opt. \textbf{25}, 1344--1349 (1986). 
 
 
 \bibitem{bide81}
A. Bideau-Mehu, Y. Guern, R. Abjean, and A. Johannin-Gilles,
J. Quant. Spectrosc. Rad. Transfer \textbf{25}, 395--402 (1981).

\bibitem{bran07}
F. Brandi, and F. Giammanco
Opt. Lett. \textbf{32}, 2327--2329 (2007).

\bibitem{bran09}
F. Brandi, F. Giammanco, W. S. Harris, T. Roche, E. Trask, and F. J. Wessel
Rev. Sci. Instrum. \textbf{80}, 113501 (2009). 

\bibitem{bran11}
F. Brandi, and F. Giammanco
Opt. Express  \textbf{19}, 25479--25487 (2011). 

\bibitem{bran13}
F Conti, M Tiberi, F Giammanco, A Diaspro, and F Brandi
 Laser Phys. Lett. \textbf{10}, 056003 (2013). 
 
 \bibitem{bran08a}
F. Brandi, P. Marsili, and F. Giammanco,
in AIP conference proceedings: Burning plasma diagnostics, \textbf{988}, 132--135 (2008).

 \bibitem{bran08}
 F. Brandi and F. Giammanco,
Opt. Lett. \textbf{33}, 2071-2073 (2008).

%
 \bibitem{malk00}
 V. Malka, C. Coulaud, J. P. Geindre, V. Lopez, Z. Najmudin, D. Neely and F. Amiranoff,
 Rev. Sci. Instrum. \textbf{71}, 2329 (2000).
 




\end{thebibliography}
\end{document}